\documentclass[%
	superscriptaddress,reprint,amsmath,amssymb,aps,pra,longbibliography]{revtex4-1}
\usepackage{dcolumn} 
\setcitestyle{numbers,super}
\usepackage{placeins} 
\usepackage{upgreek}
\usepackage{epstopdf} \usepackage{amsmath} \usepackage{physics} \usepackage{graphicx} 
\usepackage{bm} \usepackage[margin=0.7in]{geometry}
\usepackage{natbib}
\usepackage{graphicx}
\usepackage{physics}
\usepackage{newtxtext,newtxmath}

\begin{document}
\preprint{APS/123-QED}
\title{
Large-Scale Uniform Optical Focus Array Generation \\
with a Phase Spatial Light Modulator
}

\author{Donggyu Kim}
\email{donggyu@mit.edu}
\affiliation{Department of Mechanical Engineering, Massachusetts Institute of Technology, 77 Massachusetts Ave, Cambridge, MA, 02139, USA}
\affiliation{Department of Physics, Harvard University, 17 Oxford St, Cambridge, MA, 02138, USA}
\affiliation{Research Laboratory of Electronics, Massachusetts Institute of Technology, 50 Vassar St, Cambridge, MA, 02139, USA}

\author{Alexander Keesling}
\affiliation{Department of Physics, Harvard University, 17 Oxford St, Cambridge, MA, 02138, USA}

\author{Ahmed Omran} 
\affiliation{Department of Physics, Harvard University, 17 Oxford St, Cambridge, MA, 02138, USA}

\author{Harry Levine}
\affiliation{Department of Physics, Harvard University, 17 Oxford St, Cambridge, MA, 02138, USA}
    
\author{\\Hannes Bernien}
\thanks{Current address: Institute for Molecular Engineering, University of Chicago, IL 60637, USA
}
\affiliation{Department of Physics, Harvard University, 17 Oxford St, Cambridge, MA, 02138, USA}

\author{Markus Greiner}
\affiliation{Department of Physics, Harvard University, 17 Oxford St, Cambridge, MA, 02138, USA}

\author{Mikhail D. Lukin}
\affiliation{Department of Physics, Harvard University, 17 Oxford St, Cambridge, MA, 02138, USA}

\author{Dirk R. Englund}
\affiliation{Research Laboratory of Electronics, Massachusetts Institute of Technology, 50 Vassar St, Cambridge, MA, 02139, USA}
\affiliation{Department of Electrical Engineering and Computer Science, Massachusetts Institute of Technology, 50 Vassar St, Cambridge, MA, 02139, USA}

\begin{abstract}
We report a new method to generate uniform large-scale optical focus arrays (LOFAs). 
By identifying and removing undesired phase rotation in the iterative Fourier-transform algorithm (IFTA), our approach rapidly produces computer-generated holograms of highly uniform LOFAs. 
The new algorithm also shows faster compensation of system-induced LOFA intensity inhomogeneity than the conventional IFTA. 
After just three adaptive correction steps, we demonstrate LOFAs consisting of $\mathcal{O}(10^3)$ optical foci with $> 98\ \%$ intensity uniformity.
\end{abstract}

\maketitle

Uniform optical focus arrays are essential for a range of applications, including wide-field laser-scanning microscopy\cite{pang2012wide}, multifocus multiphoton microscopy\cite{bahlmann2007multifocal}, and multi-beam laser machining\cite{obata2010multi}. Recently, such focus arrays were developed as an important tool for controlling arrays of ultra-cold atoms\cite{endres2016atom, barredo2016atom, kim2016situ} for quantum computing\cite{saffman2016quantum} and quantum simulation applications\cite{bernien2017probing,PhysRevX.8.021070}. However, a challenge in these applications is the efficient production of  uniform large-scale optical focus arrays. 

A number of approaches have been developed to produce optical focus arrays, including the use of microlens arrays\cite{bahlmann2007multifocal,pang2012wide}, acousto-optic deflectors\cite{endres2016atom}, amplitude spatial light modulators (SLMs)\cite{gauthier2016direct}, and phase SLMs\cite{matsumoto2012high}. Among those, SLMs use programmable computer-generated holograms (CGHs) that enable (i) generating arbitrary focus array geometries\cite{nogrette2014single} and (ii) compensating for optical system imperfections \textit{in situ}\cite{vcivzmar2010situ,matsumoto2012high,nogrette2014single}.
Additionally, phase SLMs make more efficient use of optical power than amplitude SLMs, which necessarily attenuate light fields. To determine phase-only CGHs, iterative Fourier-transform algorithms (IFTAs), originally devised by Gerchberg and Saxton\cite{gerchberg1972practical}, are widely used\cite{di2007computer,matsumoto2014adaptive,nogrette2014single,poland2014development,tamura2016highly}. 

In this work, we demonstrate a new method for producing uniform large-scale optical focus arrays (LOFAs) with a phase SLM. By identifying and removing undesired phase rotation in the conventional IFTAs, our approach significantly reduces the number of iterations required to determine CGHs of highly uniform LOFAs consisting of $\mathcal{O}(10^3)$ optical foci. Moreover, we show that suppressing the phase rotation is essential to reliably compensate for system-induced intensity inhomogeneity, enabling rapid uniform LOFA generation in practice.

\begin{figure}[!b] 
\centering
\includegraphics[scale=0.26]{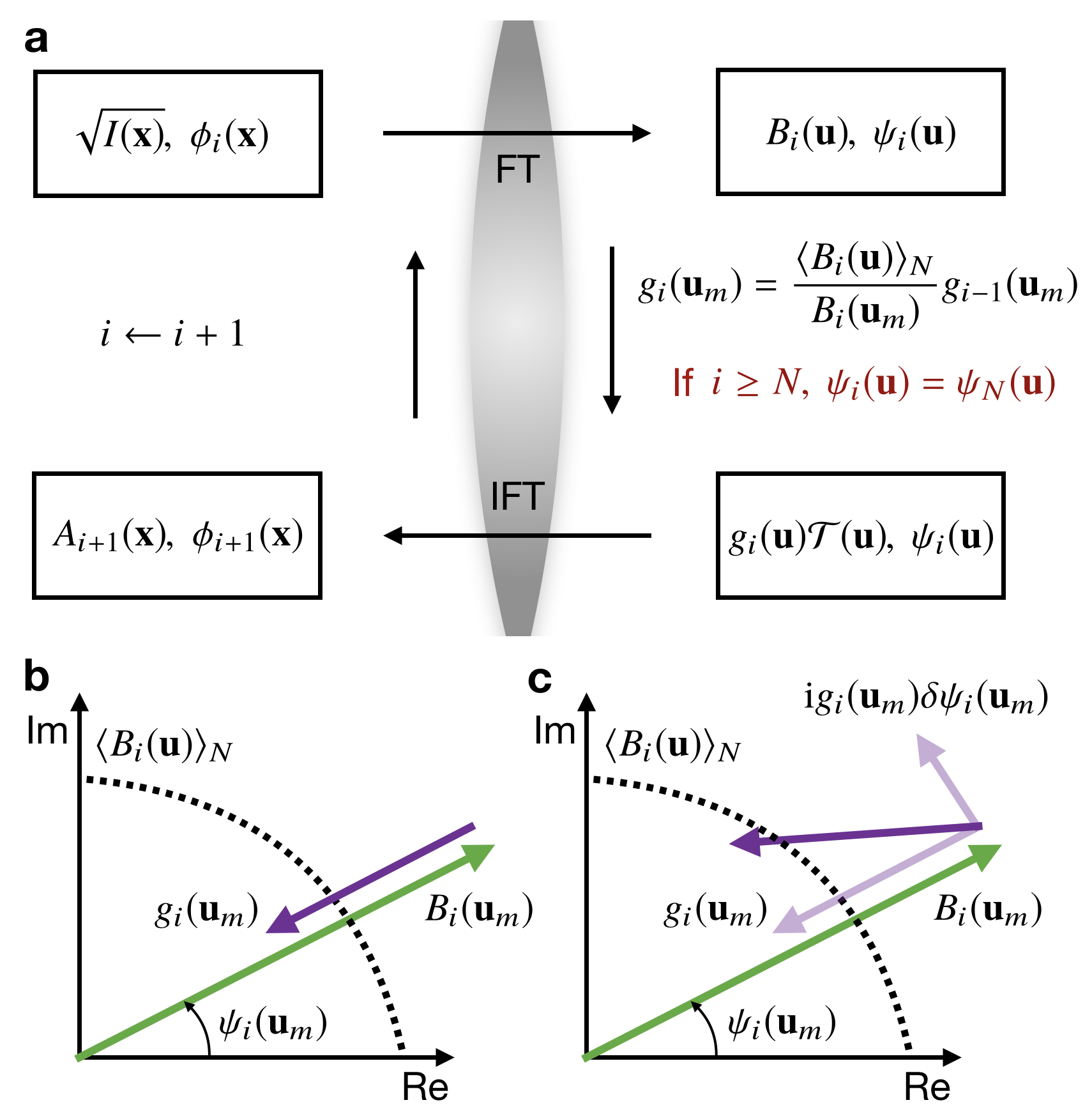}
\caption{
\textbf{Calculating computer-generated holograms (CGHs) of highly uniform large-scale optical focus arrays (LOFAs):} 
\textbf{(a)} CGHs of LOFAs are determined by iteratively updating $\phi_i (\mathbf{x})$ based on the weighted-Gerchberg Saxton algorithm. In our approach, the phase $\psi_i (\mathbf{u})$ is fixed to $\psi_N (\mathbf{u})$ for $i \geq N$. FT and IFT refer to the two-dimensional Fourier and inverse Fourier transform, respectively. $\expval{B_i(\mathbf{u} )}_N  \doteq (1/N) \sum_{m=1} ^{N} B_i (\mathbf{u}_m)$.
\textbf{(b)} $g_{i} (\mathbf{u}_m)$ updates $\phi_{i+1}(\mathbf{x})$ to reduce inhomogeneities in $B_i(\mathbf{u}_m)$ compared to $\expval{B_i(\mathbf{u} )}_N$. 
\textbf{(c)} Substituting $A_{i+1} (\mathbf{x})$ with $\sqrt{I (\mathbf{x})}$ additionally introduces inhomogeneities in $B_{i+1}(\mathbf{u}_m)$ from a phase change $\delta\psi_i (\mathbf{u}_m)$.
} 
\end{figure}

\begin{figure*}[t]
\includegraphics[scale=1]{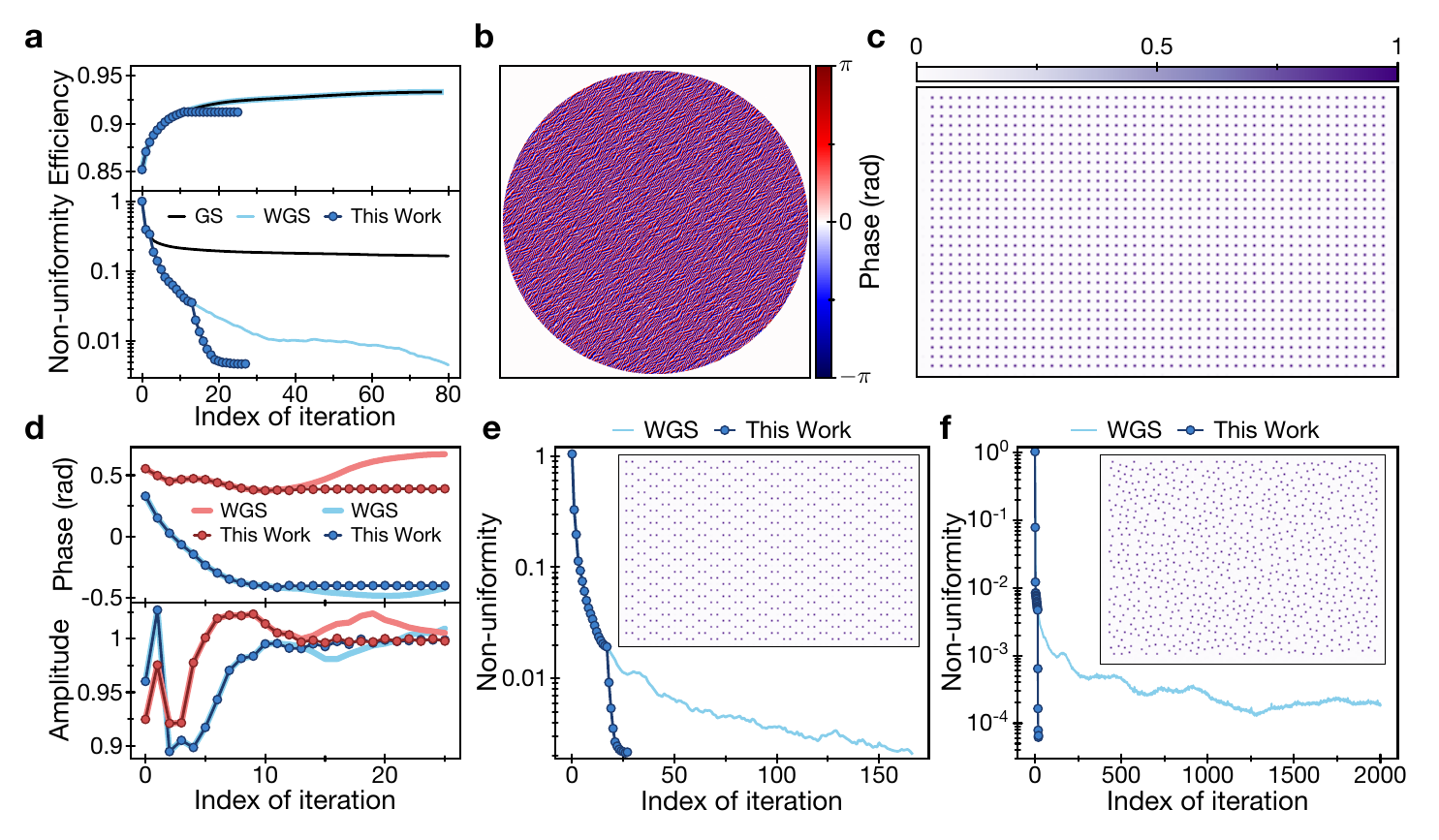}
\caption{\textbf{CGH calculation:} 
\textbf{(a)} Our approach (dark blue) rapidly achieves a desired intensity non-uniformity ($< 0.5 \ \%$, standard deviation) across 50 by 30 optical foci, compared to the conventional Gerchberg-Saxton (GS, black) and weighted-GS (WGS, light blue) algorithms (bottom). In our approach, $\psi_i (\mathbf{u})$ is fixed to $\psi_{12} (\mathbf{u})$ with a modulation efficiency of $91.2 \ \%$ (top). 
\textbf{(b)} Calculated CGH $\Phi(\mathbf{x})$ superimposed with a phase grating to spatially separate the unmodulated zeroth order in the experiment. 
\textbf{(c)} Expected LOFA at $\mathbf{u}$-plane with CGH shown in \textbf{(b)}.
\textbf{(d)} Phase (left) and amplitude (right) progress of optical focus at $\mathbf{u}_{30}$ (blue) and $\mathbf{u}_{100}$ (red) during the iterative CGH calculation.
\textbf{(e} and \textbf{f)} Comparison between our approach (dark blue) and the conventional WGS algorithm (light blue) for producing hexagonal lattice and disordered optical focus array, respectively. Insets plot expected LOFAs from iterative CGH calculation.
} 
\end{figure*}

\begin{figure}[t]
\centering
\includegraphics[scale=1]{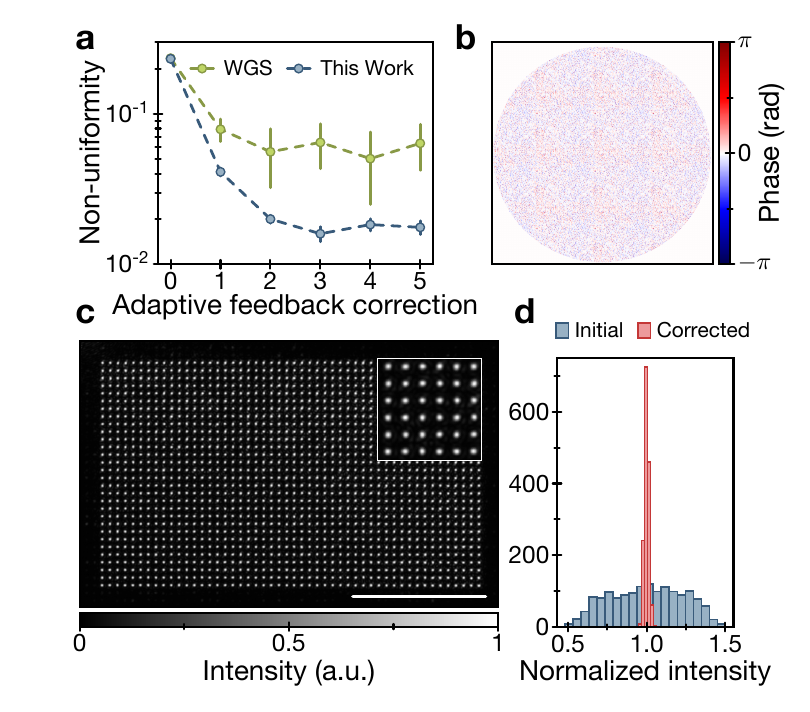}
\caption{\textbf{Uniform 50 by 30 LOFA generation with adaptive CGH correction:} 
\textbf{(a)} Adaptive correction with the conventional WGS algorithm (light green) and our phase-fixed method (dark green). Each point represents 8 ensembles of the repetitive adaptive correction, and error bars denote non-uniformity uncertainty (standard deviation) in the ensembles.
\textbf{(b)} The CGH correction $\Phi^{(3)}(\mathbf{x}) - \Phi^{(0)}(\mathbf{x})$ determined by our phase-fixed method. 
\textbf{(c)} CMOS camera raw image of 50 by 30 LOFA with $\Phi^{(3)}(\mathbf{x})$ (scale bar = 1 mm). The mean focus diameter ($1/\mathrm{e}^2$) is $15 \ \upmu \text{m}$. Inset shows a zoomed-in view. 
\textbf{(d)} Intensity histogram of optical foci. The initial intensity non-uniformity is $22 \ \%$ (standard deviation, dark green), which is reduced down to $1.4 \ \%$ (red) with the CGH correction plotted in \textbf{(b)}.
} 
\end{figure}

\section*{CGH Calculation}

We produce LOFAs at the focal plane ($\mathbf{u}$-plane) of a focusing lens. A spatially coherent light beam is reflected by a phase-SLM placed at the lens's back focal plane ($\mathbf{x}$-plane). The reflected wavefront is shaped by a CGH $\Phi (\mathbf{x})$ displayed on the SLM, forming $N$ optical foci at $\mathbf{u}_m$ ($m = 1, 2, \cdots, N$). Our method to determine $\Phi (\mathbf{x})$ is based on the weighted-Gerchberg Saxton (WGS) algorithm\cite{di2007computer} outlined in Fig. 1(a). Here, we denote the fixed incident laser intensity pattern and the target LOFA  amplitude as $I(\mathbf{x})$ and $\mathcal{T}(\mathbf{u})$, respectively.

Computing $\Phi (\mathbf{x})$ begins with the conventional WGS algorithm. The initial phase $\phi_1(\mathbf{x})$ of the algorithm consists of a random phase map uniformly distributed  from $-\pi$ to $\pi$. During the \textit{i}-th WGS iteration, the $\mathbf{u}$-plane amplitude $B_i (\mathbf{u})$ and phase $\psi_i (\mathbf{u})$ are calculated from the $\mathbf{x}$-plane amplitude $\sqrt{I(\mathbf{x})}$ and phase $\phi_i(\mathbf{x})$ by the two-dimensional Fourier transform (FT). While $\psi_i (\mathbf{u})$ is kept fixed, $B_i (\mathbf{u})$ is replaced by the target amplitude $\mathcal{T}(\mathbf{u})$ multiplied by the focus amplitude non-uniformity correction,
\begin{equation*}
g_{i} (\mathbf{u}) 
= \left[ \sum_m \frac{\expval{B_i(\mathbf{u})}_N}{B_i (\mathbf{u}_m)}\delta (\mathbf{u} - \mathbf{u}_m)\right] \times g_{i-1} (\mathbf{u}).
\end{equation*}
Here, $\expval{B_i(\mathbf{u})}_N \doteq (1/N) \sum_{m=1} ^N B_i (\mathbf{u}_m)$; $\delta(\mathbf{u})$ is the Dirac-$\delta$ function; and $g_0(\mathbf{u})=1$. This $g_i(\mathbf{u})$ updates $\phi_{i+1}(\mathbf{x})$ to compensate for irregularities in $B_{i}(\mathbf{u}_m)$ (Fig. 1(b)). Then, the corresponding $\mathbf{x}$-plane amplitude $A_{i+1}(\mathbf{x})$ and phase $\phi_{i+1} (\mathbf{x})$ are computed by the inverse two-dimensional FT. While $\phi_{i+1}(\mathbf{x})$ is used for subsequent iteration, $A_{i+1}(\mathbf{x})$ is discarded. Note that $g_i(\mathbf{u})$ has memory of the previous corrections $g_{i-1}(\mathbf{u}), g_{i-2}(\mathbf{u}), \cdots , g_{0}(\mathbf{u})$. 

This WGS iteration of $\phi_i (\mathbf{u})$ has made it possible to find a CGH $\Phi (\mathbf{x})$ of a highly uniform optical focus array\cite{di2007computer}. 
However, we note that substituting the amplitude $A_{i+1}(\mathbf{x})$ with $\sqrt{I(\mathbf{x})}$ in the WGS iteration makes $g_i (\mathbf{u})$ less effective for the following reasons: Let us consider the Fourier-transform relation $\mathcal{F}[A_{i+1}(\mathbf{x})\mathrm{e}^{\mathrm{i}\phi_{i+1}(\mathbf{x})}] = g_i(\mathbf{u}) \mathcal{T}(\mathbf{u}) \mathrm{e}^{{\mathrm{i}\psi_i(\mathbf{u})}}$ in Fig. 1(a). The amplitude substitution introduces a change $\delta\psi_i(\mathbf{u})$ in phase, which is relatively large compared to a change $\delta g_i(\mathbf{u})$ in amplitude. Thus,
\begin{align*}
\mathcal{F}[\sqrt{I(\mathbf{x})}e^{\mathrm{i}\phi_{i+1}(\mathbf{x})}] 
&\approx g_i(\mathbf{u}) \mathcal{T}(\mathbf{u}) \mathrm{e}^{{\mathrm{i}[\psi_i(\mathbf{u}) + \delta\psi_i(\mathbf{u})}]}\\
&\approx \mathcal{F}[A_{i+1}(\mathbf{x})\mathrm{e}^{\mathrm{i}\phi_{i+1}(\mathbf{x})}]
(1 + \mathrm{i}\delta\psi_i(\mathbf{u})).
\end{align*}
This phase change $\delta \psi_i (\mathbf{u})$ makes $g_i (\mathbf{u})$ less effective, since (i) $\delta \psi_i (\mathbf{u})$ additionally introduces non-uniformity in $B_{i+1}(\mathbf{u}_m)$ that is unaccounted for $g_i (\mathbf{u})$ calculation (Fig. 1(b) and (c)) and (ii) this unaccounted effect is accumulated during the iterations through the memory in $g_i (\mathbf{u})$.

We effectively remove the undesired phase change $\delta \psi_i (\mathbf{u})$. Though $\delta \psi_i(\mathbf{u})$ comes from the fixed laser intensity pattern, we note that this phase change is reflected in subsequent \textit{i}+1-th iteration  (i.e., $\psi_{i+1}(\mathbf{u}) = \psi_i(\mathbf{u}) + \delta \psi_i(\mathbf{u})$). After an initial WGS iteration of $N$ to reach a high modulation efficiency, our method removes $\delta \psi_i (\mathbf{u})$ in $i \geq N$ by fixing $\psi_i (\mathbf{u})$ to $\psi_N (\mathbf{u})$, which dramatically reduces the number of iterations to achieve a target LOFA uniformity. 
Figure 2 plots the performance of the phase-fixed method for various LOFA geometries. 
A similar phase-fixing technique was used in an earlier two-step optimization for kinoform design\cite{prongue1992optimized}.

\section*{Uniform LOFA Generation with \\ Adaptive CGH Correction}
Whereas a SLM displaying the CGH determined from the IFTAs should produce a highly uniform LOFAs, there is in practice significant inhomogeneity in the optical focus intensity, mainly due to imperfections in the optical setups and SLMs. It has been shown that such system-induced non-uniformity can be removed by adaptively correcting CGHs\cite{matsumoto2012high, matsumoto2014adaptive, nogrette2014single, poland2014development, tamura2016highly}. The generic scheme to find such CGH corrections uses the IFTAs with an adjusted target $\mathcal{T}(\mathbf{u})$ to compensate for the observed non-uniformity.

We demonstrate here that our phase-fixed method reliably and rapidly finds the adaptive CGH corrections. In our experiments, a laser beam (wavelength $\lambda$ = 795 nm) is expanded through a telescope to fill the full area of a phase SLM (X13138-02, Hamamatsu). Displaying a CGH computed from the IFTAs produces a LOFA at the focal plane of an achromatic lens (f = 250 mm), which is then imaged with a CMOS camera (DC1645C, Thorlabs). We define the initial uncorrected $\Phi(\mathbf{x})$ and its corresponding target $\mathcal{T}(\mathbf{u})$ as $\Phi^{(0)}(\mathbf{x})$ and $\mathcal{T}^{(0)}(\mathbf{u})$, respectively. 

In finding the $j$-th corrected CGH $\Phi^{(j)}(\mathbf{x})$, the CMOS camera records the LOFA image produced by $\Phi^{(j-1)}(\mathbf{x})$. Based on the recorded focus intensity $I^{(j-1)}(\mathbf{u}_m)$, the target amplitude $\mathcal{T}^{(j)}(\mathbf{u})$ is adjusted to compensate for the intensity irregularities in $I^{(j-1)}(\mathbf{u}_m)$: 
\begin{equation*}
\mathcal{T}^{(j)}(\mathbf{u})
= \left[ \sum_m \sqrt{\frac{\expval{I^{(j-1)}(\mathbf{u} )}_N}{I^{(j-1)}(\mathbf{u}_m)}}\delta (\mathbf{u} - \mathbf{u}_m)\right] \times \mathcal{T}^{(j-1)}(\mathbf{u}).
\end{equation*}
By using $\Phi^{(j-1)}(\mathbf{x})$ ($\mathcal{T}^{(j)}(\mathbf{u})$) as the initial phase (the target amplitude), $\Phi^{(j)}(\mathbf{x})$ is determined by either the conventional WGS algorithm or our phase-fixed method where the $\mathbf{u}$-plane phase $\psi_i (\mathbf{u})$ is fixed to $\psi_N(\mathbf{u})$ during the iteration for all $j$.

We first apply the adaptive CGH corrections to produce a uniform LOFA consisting of 50 by 30 optical foci (Fig. 3). The uncorrected $\Phi^{(0)}(\mathbf{x})$ (Fig. 2(b)) initially results in a LOFA with an intensity non-uniformity of 22 \% (standard deviation). Applying the repetitive CGH correction successively reduces the non-uniformity as plotted in Fig. 3(a). Compared to the correction using the conventional WGS algorithm (light green), our phase-fixed method (dark green) performs more reliable and rapid CGH correction, achieving $1.59 \pm 0.18 \ \%$ non-uniformity only with three correction steps. The correction with the WGS algorithm achieves $6.47 \pm 2.14 \ \%$ with the same number of corrections.

\begin{figure}[b] 
\centering
\includegraphics[scale=0.26]{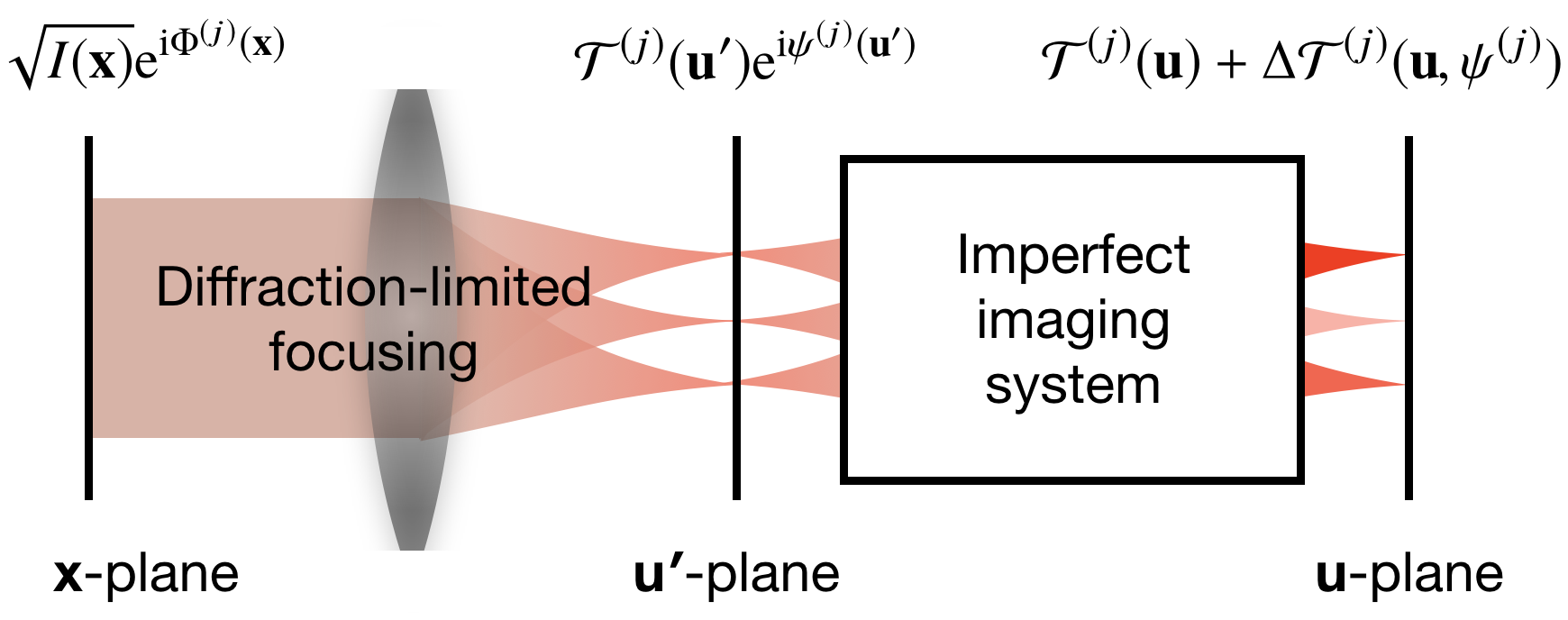}
\caption{
\textbf{Three-plane model:} 
Ideal SLM and diffraction-limited focusing lens form the target LOFA ($\mathcal{T}^{(j)}(\mathbf{u}')\mathrm{e}^{\mathrm{i}\psi^{(j)}(\mathbf{u}')}$) at the lens' focal plane. All the system imperfections are lumped into a virtual imaging system, introducing the focus amplitude non-uniformity $\Delta \mathcal{T}^{(j)}(\mathbf{u}, \psi^{(j)}(\mathbf{u}'))$ at  the $\mathbf{u}$-plane.
} 
\end{figure}

\begin{figure*}[t] 
\centering
\includegraphics[scale=1]{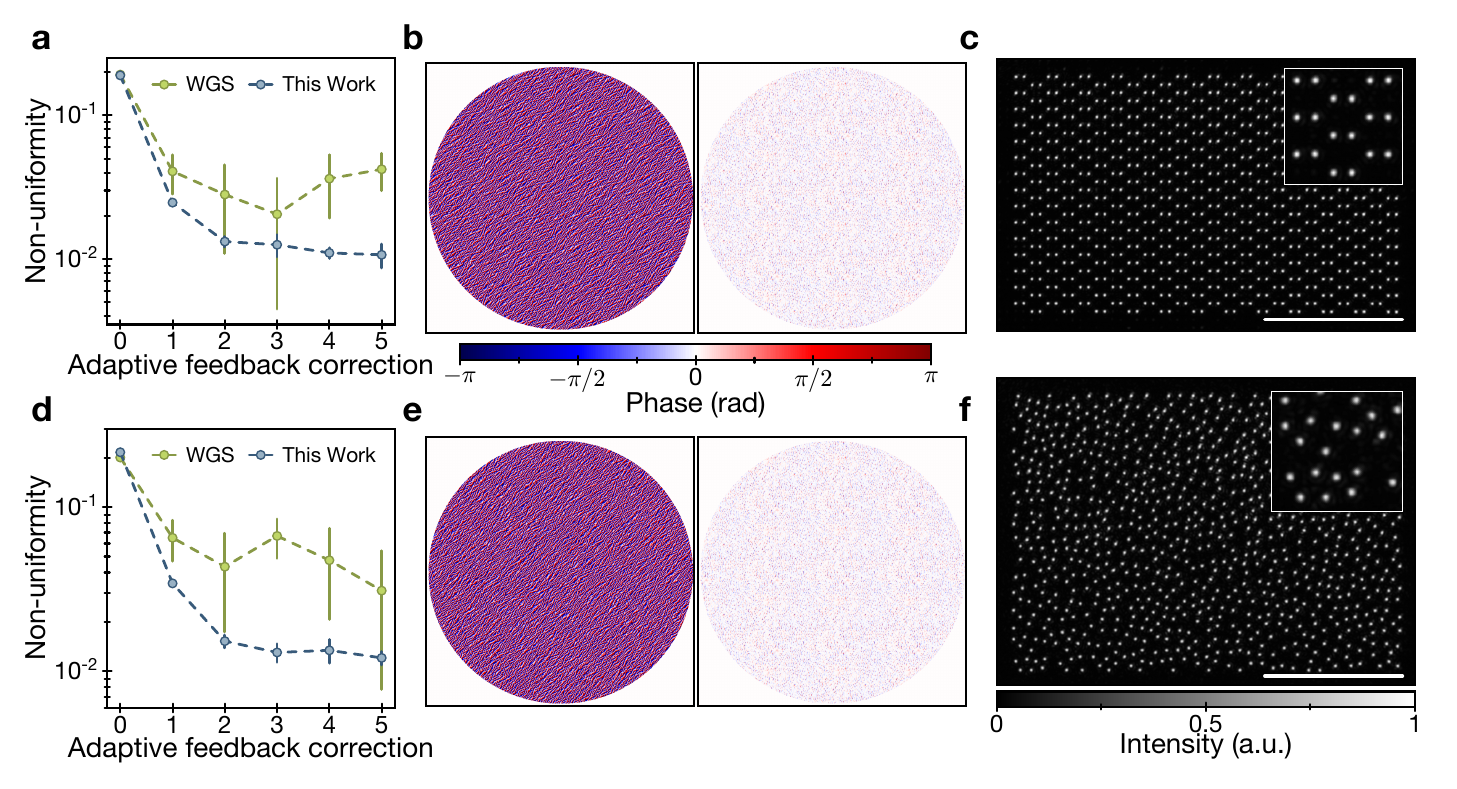}
\caption{\textbf{Hexagonal lattice and disordered LOFA generation:} 
\textbf{(a)} Adaptive correction with the conventional WGS algorithm (light green) and our phase-fixed method (dark green). Each point represents 8 ensembles of the repetitive adaptive correction, and error bars denote non-uniformity uncertainty (standard deviation) in the ensembles.
\textbf{(b)} (Left) uncorrected CGH $\Phi^{(0)}(\mathbf{x})$ superimposed with a phase grating to spatially separate the unmodulated zeroth order. (Right) the CGH correction $\Phi^{(5)}(\mathbf{x})-\Phi^{(0)}(\mathbf{x})$ determined by our phase-fixed method. 
\textbf{(c)} CMOS camera raw image with $\Phi^{(5)}(\mathbf{x})$ (scale bar = 1 mm). Inset shows a zoomed-in view.
\textbf{(d)-(f)} Same as \textbf{(a)-(c)} for disordered LOFA geometry.
} 
\end{figure*}

The reliable and rapid correction due to our phase-fixed method can be understood from the three-plane model illustrated in Fig. 4. Here, the SLM and focusing lens are assumed to be ideal, thus $\Phi^{(j)}(\mathbf{x})$ forms the target $\mathcal{T}^{(j)}(\mathbf{u}') \exp [\mathrm{i}\psi^{(j)}(\mathbf{u}')]$ on the intermediate $\mathbf{u}'$-plane. $\psi^{(j)}(\mathbf{u}')$ is specified by running the \textit{j}-th correction IFTA. All the system-imperfections are lumped into a virtual imaging system that maps $\mathbf{u}'$- to $\mathbf{u}$-plane with a measured irregularity $\Delta \mathcal{T}^{(j)}(\mathbf{u})$. In this model, we ignore the non-uniformity expected from the IFTA ($\mathcal{O}(10^{-3})$), which is much smaller than one from the measurement ($\mathcal{O}(10^{-2})$). Note that $\Delta \mathcal{T}^{(j)} (\mathbf{u})$ clearly depends on $\psi^{(j)}(\mathbf{u}')$ in this configuration. 

Let us first consider adaptive CGH corrections using the conventional WGS algorithm. When $\Phi^{(j)}(\mathbf{x})$ is computed from $\Phi^{(j-1)}(\mathbf{x})$, $\psi^{(j)}(\mathbf{u}')$ rotates from $\psi^{(j-1)} (\mathbf{u}')$. 
This phase rotation introduces an additional non-uniformity in $\Delta \mathcal{T}^{(j)}(\mathbf{u})$, which is unaccounted when $\mathcal{T}^{(j)} (\mathbf{u})$ is adjusted from $\Delta \mathcal{T}^{(j-1)} (\mathbf{u})$. 
Moreover, CMOS noise in $\Delta \mathcal{T}^{(j-1)} (\mathbf{u})$ also contributes to the phase rotation and adds the subsequent non-uniformity noise in $\Delta \mathcal{T}^{(j)}(\mathbf{u})$. 
By contrast, our phase-fixed method to compute $\Phi^{(j)}(\mathbf{x})$ is free from such phase rotations, enabling reliable and rapid correction of $\Delta \mathcal{T}^{(j-1)} (\mathbf{u})$.
 
The use of SLMs allows to generate arbitrary LOFA geometries such as a hexagonal lattice and disordered geometries (Fig. 5). These geometries in particular are useful for quantum many-body physics simulation\cite{gorg2018enhancement} and quantum optimization\cite{pichler2018quantum}. We demonstrate an uniform LOFA of the hexagonal lattice (disordered) geometry consisting of 720 (819) foci. By applying our phase-fixed method, we achieve a LOFA non-uniformity of $1.1 \pm  0.20 \ \%$ ($1.2 \pm 0.11 \ \%$) with the adaptively corrected CGH $\Phi^{(5)}(\mathbf{x})$ (Fig. 5(b) and (e)). Both cases show greater reliability and convergence of the phase-fixed method compared to the conventional WGS algorithm to counteract the system-induced non-uniformity.

\vspace{5mm}

\section*{Conclusion}
We present a new method to generate uniform large-scale optical focus arrays (LOFAs) using a phase spatial light modulator. First, we identified and avoided the undesired phase rotation in the conventional weighted-Gerchberg Saxton (WGS) algorithm. As a result, our approach significantly reduces the number of iterations to determine CGHs of highly uniform LOFAs compared to the conventional WGS algorithm. Next, we experimentally showed that this phase-fixed approach allows us to more reliably and rapidly counteract optical system imperfections that degrade LOFA's intensity uniformity. With only three adaptive correction steps, our approach produces arbitrary LOFA geometries consisting of $\mathcal{O}(10^3)$ foci with a uniformity $> 98 \  \%$. This new algorithm that rapidly produces uniform LOFAs should benefit a range of applications including optical microscopy, optical information processing, and quantum control for atoms and solid-state quantum emitters. 

\vspace{3mm}

\section*{Acknowledgments}
This work was supported by NSF, CUA, and Vannevar Bush Faculty Fellowship. D.K. was supported in part by the AFOSR MURI for Optimal Measurements for Scalable Quantum Technologies (FA9550-14-1-0052) and by the Army Research Office Multidisciplinary University Research Initiative (ARO MURI) biological transduction program.


%

\end{document}